\documentclass[%
reprint,
showpacs,
showkeys,
amsmath,amssymb,
aps,
prd
]{revtex4-2} 

\usepackage{graphicx}
\usepackage{xcolor}
\usepackage{pgffor}
\usepackage{capt-of}
\usepackage{soul}
\usepackage{lipsum}
\usepackage{enumerate}
\usepackage{tikz}
\usepackage{rotating}
\usepackage{url}

\setcitestyle{maxauthors=3}

\usepackage{marginnote}

\usetikzlibrary{decorations.pathmorphing}
\usetikzlibrary{decorations.text}
\usetikzlibrary{positioning}
\usetikzlibrary{patterns}

\def\beq{\begin{equation}}
\def\eeq{\end{equation}}
\def\bea{\begin{eqnarray}}
\def\eea{\end{eqnarray}}

\let\rho=\varrho

\begin{document}

\frenchspacing

\title[] {Phase transitions in self-gravitating objects and black holes: Common properties and interpretations}

\author{Martin Blaschke}
\email{martin.blaschke@physics.slu.cz}
\author{Petr Slan{\' y}}
\email{petr.slany@physics.slu.cz}
\affiliation{%
	Institute of Physics and Research Centre of Theoretical Physics and Astrophysics, Faculty of Philosophy and Science,
	Silesian University in Opava,\\
	Bezru\v{c}ovo n\'am\v est\'i~13, CZ-746\,01 Opava, Czech Republic}


\begin{abstract}
We develop a simple analytic two-component model that links the thermal evolution of self-gravitating systems to black-hole thermodynamics. The model consists of a monatomic nonrelativistic ideal gas and an effective barotropic support calibrated by the specific heat $\hat c_V^{\rm J}(\Theta)$ of a J\"uttner gas. From the partition function we obtain an exact modified Bessel expression for $\hat c_V^{\rm J}(\Theta)$ and supply compact Pad\'e approximants with good accuracy across all temperatures. These inputs yield a closed virial-capacity formula, which we use to determine where the heat capacity changes sign as the support pressure becomes competitive with the thermal pressure. We then contrast this behavior with Schwarzschild anti-de Sitter black holes. Despite the disparate microphysics, both systems exhibit a switch in heat-capacity behavior governed by competing pressure contributions. The surrogate model provides compact analytic insight into stability thresholds in self-gravitating systems and their black-hole analogues.
\end{abstract}

\maketitle 

\begin{center}
\footnotesize
\copyright\ 2026 American Physical Society. This is the accepted manuscript of the following article:\\
M. Blaschke and P. Slan\'y, ``Phase transitions in self-gravitating objects and black holes: Common properties and interpretations,''
\textit{Phys. Rev. D} \textbf{114}, 104060 (2026).\\
The final published version is available at \url{https://doi.org/10.1103/ynrz-byz5}.
This manuscript version is posted with permission for non-commercial scholarly use.
\end{center}
\vspace{0.5em}


\section{Introduction}

Black holes (BHs) have transformed our understanding of spacetime and general relativity \cite{2018grav}. The detection of gravitational waves by LIGO in 2015 \citep{2016PhL} offered direct evidence of BH mergers \cite{2017PhRvL,2022Abb:prd:}, marking a new age in observational astronomy, and the Event Horizon Telescope subsequently imaged the shadow of a supermassive BH \cite{2023arXi}. Beyond their observational relevance, BHs occupy a unique place in physics due to their extreme compactness and their role as a testing ground for quantum aspects of gravity, including the information paradox (for review see, e.g., \cite{Suv:2020:}) and BH thermodynamics (for review see, e.g., \cite{Car:2014:,Wit:2024:}). 

Self-gravitating systems display thermodynamic behavior with no analogue in short-range systems, most notably an \emph{effective negative heat capacity} during gravitational contraction: as energy is lost, the temperature rises until an additional pressure source---typically electron degeneracy or nuclear burning---stabilizes the core \cite{Bell:Bell:1977:}. This textbook mechanism suggests a simple conceptual picture with three stages: (i) a thermal-pressure dominated regime (classical gas, often unstable in the effective thermodynamic sense), (ii) a competition regime where an additional support becomes comparable to thermal pressure, and (iii) a support-dominated regime in which the system can be thermally stabilized and cool.

Black holes lie at the extreme end of this story. In the extended phase-space approach to BH thermodynamics, the cosmological constant $\Lambda$ is interpreted as vacuum pressure, while the black-hole mass $M$ is thermodynamically interpreted as enthalpy \cite{Kas:etal:2009:,Kub:Man:Teo:2017:}. This framework introduces thermodynamic pressure $p$ and volume $V$ and allows the analysis of BH heat capacities, including Davies critical points that organize small/large BH phases \cite{Dav:1977:,Henn:Tei:1984:,Tei:1985:,Alt:Kub:Mann:2013:,Kub:Man:2015:,Mann:2025:}.

Recent work has reinterpreted these discontinuities in terms of a balance of horizon pressures (via Newman--Penrose scalars), where the divergence occurs when a thermal (Komar/equipartition) pressure matches the appropriate nonthermal contributions at the horizon \cite{Bar:2024:}. This suggests that heat-capacity behavior in BHs and in self-gravitating systems can be phrased in a common language of competing pressure supports.

\subsection{Step-by-step construction of the surrogate model.}

Motivated by this picture and Ref.~\cite{Bar:2024:}, our goal is to build an analytic \emph{surrogate} that captures: (a) a pressure contribution linear in $T$ (thermal regime), (b) an additional support that stiffens under compression and can become nearly temperature independent (support regime), and (c) an explicit criterion for the switch of the effective heat capacity along gravitational contraction. To this end we model a self-gravitating system as a \emph{two-component} system in thermal equilibrium:
\begin{itemize}
\item[(i)] a \emph{monatomic nonrelativistic ideal-gas} thermal component, representing baryons/ions (e.g. protons) that remain well approximated as nonrelativistic in a wide range of stellar-core conditions; 

\item[(ii)] an \emph{effective barotropic (polytropic) support} $p_{\rm eff}=\mathcal{K}\varrho^{\,n}$, emulating degeneracy-like pressure support (e.g. relativistic electrons) while retaining analytic tractability. Here $\mathcal{K}$ sets the scale of the support term, while the constant-index $n$
case is later promoted to an effective temperature-dependent index $n(T)$. 
\end{itemize}
The corresponding caloric equation of state (CEOS) is taken in the form
\begin{equation}
U=\frac{3}{2}N k_{\mathrm B} T+\frac{\mathcal{K}}{n-1}\,\varrho^{\,n}V,
\label{eq:intro_U}
\end{equation}
where the factor $1/(n-1)$ ensures thermodynamic consistency with a polytropic response.

The key idea is to promote the polytropic index $n$ from a constant to a temperature-dependent effective index calibrated by relativistic kinematics,
\begin{equation}
n(T)=1+\frac{1}{\hat c_V^{\rm J}(\Theta)}\, ,
\label{eq:intro_n_def}
\end{equation}
where the dimensionless relativistic parameter is defined as
\begin{equation}
	\Theta = \frac{k_{\mathrm B}T}{mc^2}\, .
	\label{relpar}
\end{equation}

Therefore the relativistic kinematics enters the picture through the J\"uttner specific heat $\hat c_V^{\rm J}(\Theta)$, yielding the correct limits
$n\to5/3$ (nonrelativistic) and $n\to4/3$ (ultrarelativistic).
In the pressure-competition viewpoint \cite{Bar:2024:}, this provides a minimal analytic realization of a thermal term competing with a nonthermal support term, and the stability change is encoded in the sign switch of the effective (virial-path) heat capacity.

From the single-particle partition function we derive an exact expression for $\hat c_V^{\rm J}(\Theta)$ in terms of modified Bessel functions $K_\nu$, and we provide compact Pad\'e approximants that track the exact result with good accuracy over the entire temperature range. These inputs yield a virial capacity formula
\begin{equation}
C_\mathrm{vir} \;=\; \frac{C_V}{\,1-\dfrac{2T}{p}\left(\dfrac{\partial p}{\partial T}\right)_{V}\,},
\label{eq:intro_Cvir}
\end{equation}
which we use to map the sign of $C_\mathrm{vir}$ across $(\varrho,\Theta)$. The heat capacity changes sign as the support pressure becomes competitive with the thermal pressure, providing a sharp criterion for stabilization against collapse.

In a microscopic degeneracy model the polytropic prefactor $\mathcal{K}$ also depends on composition and relativistic corrections. In the present surrogate model we keep $\mathcal{K}$ constant to isolate the effect of the temperature-dependent stiffness encoded in $n(T)$; $\mathcal{K}$ therefore sets the overall scale of the support term.

We then compare this behavior with Schwarzschild anti-de Sitter (AdS) thermodynamics. Because the BH enthalpy exhibits a degenerate Legendre structure, one finds $C_V=0$ and the natural comparison is done with $C_p$ instead \footnote{Since the BH thermodynamics description does not involve a particle-number degree of freedom, it is natural to compare it with thermodynamic systems at vanishing chemical potential $\mu=0$; for a pedagogical discussion of $\mu=0$ thermodynamics see \cite{Blaschke_2026}.}. The \emph{Davies point} $p=\pi_T$ separates the small/large BH phases via a divergence of $C_p$ \cite{Dav:1977:,Haw:Pag:1983:,Kub:Man:Teo:2017:}. In this side-by-side view, both systems exhibit a switch in heat capacity behavior governed by a competition between a collapse driving term and a stabilizing pressure \cite{Bar:2024:}.

\subsection{Scope and limitations.}

The model is a thermodynamic surrogate designed to interpolate smoothly between relativistic regimes in closed form. Its thermal calibration describes relativistic kinetic effects, while a microscopic $T\!\to\!0$ Fermi--Dirac treatment lies beyond its scope. We apply the Newtonian/quasi-static virial analysis to self-gravitating systems in the weak-field regime, whereas the BH part is a thermodynamic comparison in the extended phase space of static, uncharged AdS spacetimes. Strongly compact systems would require a relativistic hydrostatic treatment. Within these bounds, the model provides a compact analytic view of heat-capacity sign changes in self-gravitating systems and their analogy with BH thermodynamics.

\section{Two-component model}

We work in the Newtonian/quasi-static virial framework appropriate for self-gravitating systems. We denote the global nongravitational internal energy of the system by $U$ and its global Newtonian gravitational potential energy by $\Omega$ (both understood as volume integrals over the configuration). The `virial path' denotes a sequence of near-equilibrium states satisfying the standard Newtonian virial relation.

We consider the total energy of the self-gravitating system as
\begin{equation}
	E=U+\Omega\, .
\end{equation}

For a nonrelativistic monatomic ideal gas with negligible surface pressure,
the hydrostatic virial theorem reduces to the familiar form
$\Omega=-2U$. More generally, the virial relation is written in terms of
the pressure integral $3\int p\,dV$, as discussed in Appendix~3.
Within the global thermodynamic approximation used below, this relation
is represented by an effective virial balance between $\Omega$ and the
nongravitational internal energy.

We model the self-gravitating system as a monatomic nonrelativistic ideal-gas thermal component together with an effective barotropic support component. Its caloric equation of state is
\begin{equation}\label{CEOS}
	U = \frac{3}{2}N k_\mathrm{B} T
	+ \frac{\mathcal{K}}{n(T)-1}\rho^{n(T)}V\, .
\end{equation}
Here $\rho=M/(V\rho_0)$ is the dimensionless mean-density variable of the system. At fixed $T$ and fixed total mass $M$, the index $n(T)$ is constant under the volume derivative, and
\begin{equation}
	\left(\frac{\partial U}{\partial V}\right)_T
	=
	-\mathcal K\rho^{n(T)} .
	\label{eq:dUdV-surrogate}
\end{equation}

The CEOS specifies $U(T,V)$. To complete the thermodynamic specification, we use the pressure equation of state
\begin{equation}
	p(T,V)=p_{\rm th}(T,V)+p_{\rm supp}(T,V),
	\label{EOS}
\end{equation}
with
\begin{equation}
	p_{\rm th}V=Nk_{\rm B}T .
\end{equation}
The pressure relation is constrained by the thermodynamic identity for the internal pressure,
\begin{equation}
	\pi_T
	\equiv
	\left(\frac{\partial U}{\partial V}\right)_T
	=
	T\left(\frac{\partial p}{\partial T}\right)_V-p\, .
	\label{eq:internal-pressure-model}
\end{equation}
For the ideal-gas thermal contribution,
\begin{equation}
	T\left(\frac{\partial p_{\rm th}}{\partial T}\right)_V-p_{\rm th}=0 .
\end{equation}
Using Eq.~\eqref{eq:dUdV-surrogate}, the support contribution therefore satisfies
\begin{equation}\label{rov9}
	\left(\frac{\partial}{\partial T}\frac{p_{\rm supp}}{T}\right)_V
	=
	-\frac{\mathcal K\rho^{n(T)}}{T^2}\, .
\end{equation}
The general solution can be written as
\begin{equation}
	p_{\rm supp}(T,V)
	=
	\mathcal K T
	\int_T^\infty
	\frac{\rho^{n(\tau)}}{\tau^2}\,d\tau
	+T f(V),
\end{equation}
where $f(V)$ is the homogeneous integration term. For constant $n$, the integral gives $\mathcal K\rho^n$. Requiring the standard polytropic result $p_{\rm supp}=\mathcal K\rho^n$ in this limit fixes $f(V)=0$. The support pressure used throughout the model is therefore
\begin{equation}
	p_{\rm supp}(T,V)
	=
	\mathcal K T
	\int_T^\infty
	\frac{\rho^{n(\tau)}}{\tau^2}\,d\tau .
	\label{eq:support-pressure-closure}
\end{equation}

Both contributions share the same $T$, $V$, $\rho$, and Newtonian gravitational field. This decomposition separates the thermal pressure from the additional stabilizing support and allows their competition to be tracked through the virial heat capacity.

The temperature dependence of the effective polytropic index is fixed by a relativistic kinetic calibration. We use the dimensionless heat capacity of the classical J\"uttner gas,
\begin{equation}
	\hat c_V^{\rm J}(\Theta)
	=
	\frac{C_V^{\rm J}(\Theta)}{Nk_{\rm B}}\,  .
	\label{eq:juttner-dimensionless-cv}
\end{equation}

For the J\"uttner gas, Mayer's relation gives the adiabatic index
\begin{equation}
	\gamma_{\rm J}(\Theta)
	=
	\frac{C_P^{\rm J}}{C_V^{\rm J}}
	=
	1+\frac{1}{\hat c_V^{\rm J}(\Theta)} .
\end{equation}
We calibrate the effective polytropic index by setting
\begin{equation}
	n(T)\equiv\gamma_{\rm J}(\Theta)
	=
	1+\frac{1}{\hat c_V^{\rm J}(\Theta)} .
	\label{eq:n-calibration}
\end{equation}
Thus relativistic kinematics modifies the effective stiffness of the support term, interpolating between the familiar nonrelativistic and ultrarelativistic polytropic indices,
\begin{eqnarray}
\label{AsI}
n(T)&\rightarrow& \frac{5}{3} \quad  \left(T\rightarrow 0\right)\, ,\nonumber \\ 
n(T)&\rightarrow&\frac{4}{3} \quad  \left(T\rightarrow  \infty\right)\, .
\end{eqnarray}
The heat capacity of the complete surrogate model is obtained independently from its CEOS,
$C_V^{\rm model}=(\partial U/\partial T)_V$. The explicit temperature dependence of the J\"uttner calibration function is derived below.

The model describes a hybrid system in which relativistic effects are represented thermodynamically through a variable heat capacity, allowing the support sector to approach the degeneracy-like EOS $p\sim\rho^{4/3}$ in the high-energy limit. A microscopic fully degenerate Fermi gas at low temperature lies outside this surrogate description. The temperature $T$ therefore serves as an effective parameter tracking the relativistic softening of the support sector in the surrogate model.

Furthermore, both components are taken to contain the same number of particles $N$. Therefore, this  model can describe an object in which, e.g., only free relativistic electrons are degenerate while protons are still non-relativistic and classical particles. Hence the model could be suitable for, e.g., white dwarfs.   

Equation~\eqref{eq:support-pressure-closure} is used throughout the interpolation region. In the asymptotic regimes, where $n(T)$ approaches a constant, it reduces to the standard polytropic support pressure used below.

\section{Exact and Padé approximations for the heat capacity}

\begin{widetext}
\begin{minipage}{0.48\linewidth}
	\begin{tikzpicture}[thick, scale=1]
				\draw[-latex,white,postaction={decorate},decoration={text along path,
text={exact},text align=center}]
(-3.75,-0.7) arc [start angle=150,end angle=75,radius=4.9];
		\node[inner sep=1pt] (img1) at (-0.5,0) {\includegraphics[width=0.9\textwidth]{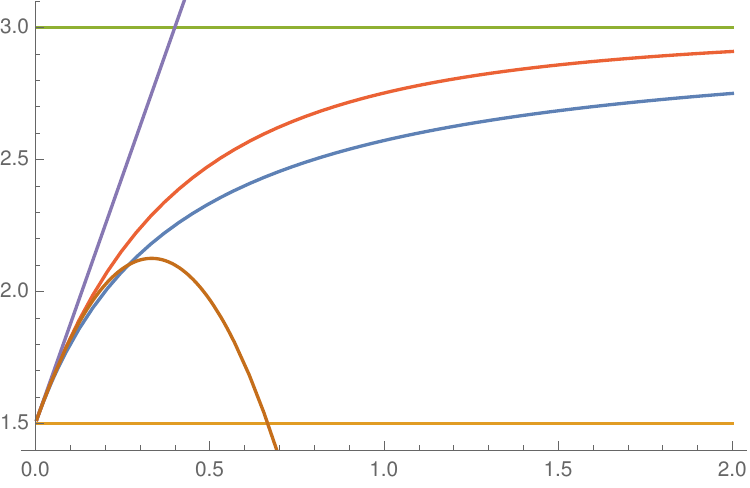}};
		\node at (0,-2.7) {\large $\Theta$};
		\node at (-4.7,0) {\large $\hat{c}_V$};
		\node at (2.1,-1.65) {nonrelativistic};
		\node at (2.0,2.45) {ultra relativistic};
		\node[rotate=6] at (2.1,1.25) {Padé, 1. order};
		\node[rotate=70] at (-3.3,0.7) {Taylor, 1. order};
		\node[rotate=-65] at (-1.85,-0.75) {2. order};	
		\draw[dashed,gray] (0.7,1.7) rectangle (2,2);

	\end{tikzpicture}
	\captionof{figure}{\label{S1}  Comparison of the capacities $\hat{c}_V$ at five different levels of accuracy. Matched
asymptotic expansions and exact Bessel form  merge with ultra relativistic case at $\Theta\rightarrow \infty$.}
\end{minipage}
\hfill
\begin{minipage}{0.48\linewidth}
	\begin{tikzpicture}[thick, scale=1]
		
		\node[inner sep=1pt] (img2) at (-0.5,0) {\includegraphics[width=0.9\textwidth]{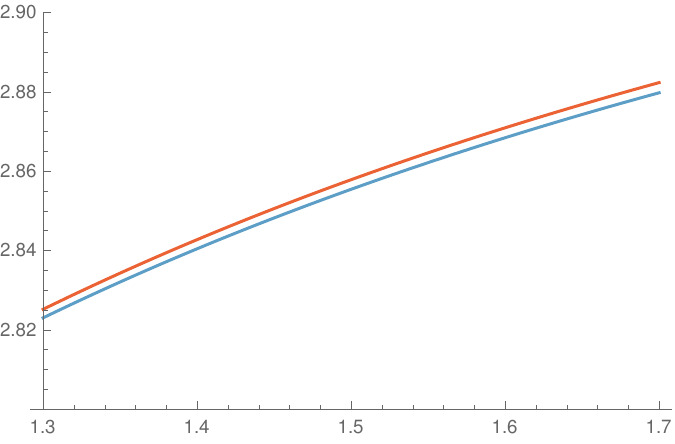}};
		\node at (0,-2.7) {\large $\Theta$};
		\node at (-4.7,0) {\large $\hat{c}_V$};
		\node[rotate=18] at (1,1.2) {exact};	
		\node[rotate=18] at (1,0.6) {Padé 2. order};	
	\end{tikzpicture}
	\captionof{figure}{\label{S11} Zoom of Fig.~\ref{S1}. showing how accurate the second-order Padé approximant already is.\\ }
\end{minipage}
\end{widetext}

In this section we derive the relativistic J\"uttner specific heat $\hat{c}_V^{\rm J}(T)$ from the single-particle partition function. To keep the notation light within this section, we suppress the superscript ${\rm J}$ on $\hat c_V$. We then develop accurate matched asymptotics using Padé approximants to produce simple closed-form formulas with solid accuracy across all temperatures.

The partition function describing one free relativistic classical (spinless) particle is
\begin{eqnarray}\label{OPF}
	z_1 = \frac{4\pi V}{\left(2\pi\hbar\right)^3}\left(\int\limits_{0}^{\infty}p^2\mathrm{e}^{-\beta\left(\sqrt{p^2c^2+m^2c^4}-mc^2\right)}dp\right)\, . 
\end{eqnarray}	
Here we have standardly subtracted the rest energy $mc^2$
 in the Boltzmann factor so that the resulting thermodynamic potentials describe the thermal contribution above rest mass; this amounts to a constant shift of the energy zero and does not affect response functions.
 
The integral (\ref{OPF}) can be solved with the aid of the modified Bessel functions of the second kind $K_{\nu}(x)$ with the result (see Appendix 1.)
\begin{equation}
	z_1 =a V \mathrm{e}^{1/\Theta}\Theta K_2(1/\Theta)\, ,
\end{equation} 
where the constant $a$ is
\begin{equation}
a =\frac{m^3c^3}{2\pi^2\hbar^3}\, .
\end{equation}

Following standard statistical methods (see Appendix~2 for more details) we easily find relativistic versions of the thermal capacity in the form:
\begin{eqnarray}\label{RelC}
		\hat c_V
		= 3 + \frac{1}{\Theta^{2}}
		- \frac{3}{\Theta}\,\frac{K_{1}}{K_{2}}
		- \frac{1}{\Theta^{2}}\!\left(\frac{K_{1}}{K_{2}}\right)^{\!2}\, ,
\end{eqnarray}
where $K_\nu\equiv K_\nu(1/\Theta)$ are Bessel functions with argument $1/\Theta$ (from now on, every Bessel function in the manuscript is to be understood as a function of $1/\Theta$).

Electron temperatures in the core of the most energetic stars reach $\Theta \sim 0.2$. This is the main reason why we have chosen the CEOS to be composed of two parts. First part represents protons behaving like non-relativistic ideal gas, second part describes degenerate electrons approximated by relativistic polytropic response.

\subsection{Matched asymptotic expansions via the Padé method}

In this section we will provide asymptotic relation for the function $\hat{c}_V$ in order to simplify complicated relations containing Bessel functions without significant loss in accuracy.  

Since the Taylor expansion is also an asymptotic one, for small temperatures we easily find asymptotic relation 
\begin{eqnarray}
	\hat{c}_{V} &\sim& \frac{3}{2}+\frac{15\Theta}{4} -\frac{45 \Theta^2}{8}+\cdots\, ,\quad (\Theta\rightarrow 0)\, .
\end{eqnarray}
Same logic can be applied to large temperatures, with the result 
\begin{eqnarray}
	\hat{c}_{V} &\sim& 3-\frac{1}{2\Theta^2}+\cdots\, ,\quad (\Theta\rightarrow \infty)\, .
\end{eqnarray}

Next we would like to link these two local results to find global approximation called matched asymptotic. To achieve this goal the Padé method is an excellent tool, because closed form of the $\hat{c}_V$ given in Eq.~\eqref{RelC} is in the form of Bessel functions rationals. Padé method can therefore mimic the singularities of Bessel functions and the resulting asymptotic analysis can be remarkably precise.      
 
Resulting 2-point $(\Theta\rightarrow 0$ and $\infty)$ Padé approximants of the first and second order are:
\begin{eqnarray}
	\hat{c}_{V\mathrm{P1}} &\sim& \frac{3}{2}+\frac{15\Theta}{4+10\Theta}\, ,\\
	\hat{c}_{V\mathrm{P2}}  &\sim& \frac{3}{2}+\frac{15\Theta+15\Theta^2}{4+10\Theta+10\Theta^2}\, .
\end{eqnarray}

In Fig. \ref{S1} we see that the 1. order approximation seems to be quite good, but the integrated relative deviation, i.e. 
\begin{equation}
	\bar{\delta}_1=\int\limits_{0}^\infty \frac{\hat{c}_{V\mathrm{P1}}-\hat{c}_V}{\hat{c}_{V}} d\Theta\, 
\end{equation}
is actually infinite. It is because matched asymptotic $\hat{c}_{V\mathrm{P1}}$ converges to correct value 3 as $\Theta\rightarrow\infty$, but too slowly. 

\begin{figure}[t]
	\begin{tikzpicture}[thick, scale=1]
		
		\node[inner sep=1pt] (img1) at (-0.5,0) {\includegraphics[width=0.45\textwidth]{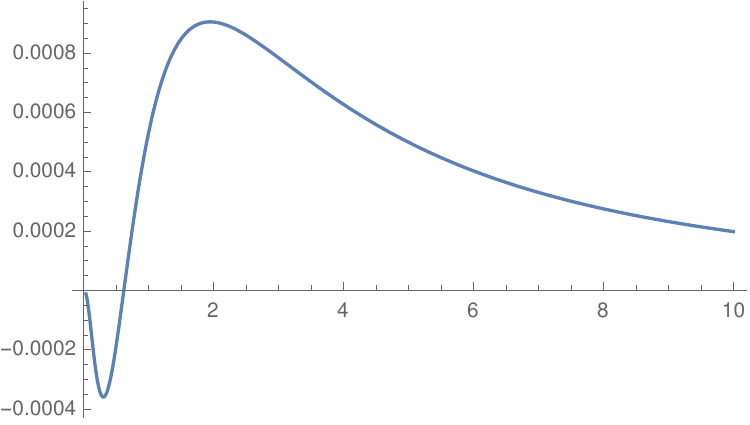}};
		\node at (0,-2.7) {\large $\Theta$};
		\node at (-4.7,0) {\large $\delta_{2}$};
	\end{tikzpicture}
	\caption{\label{S2} Relative error of 2. order Padé as a function of the relativistic parameter $\Theta$. Its integrated relative deviation is $\bar{\delta}_{2}=0.0073$.}	
\end{figure}

The Padé approximant of the 2. order is (as it is typical for matched asymptotic) already so precise that it cannot be distinguished from the exact relation via Bessel functions in Fig. \ref{S1}. Therefore, we have provided a zoomed view in Fig. \ref{S11}. The integrated relative deviation is $\bar{\delta}_{2}=0.0073$ (see Fig. \ref{S2}).

Interestingly, the 3. order of Padé diverges and to achieve more accuracy we would have to deviate from pure Padé methods and introduce logarithmic functions into the asymptotic expansion.   

Since the index $n$ does not contain any transcendental functions of $\hat{c}_V$, its asymptotic relation can be obtained via simple substitution $\hat{c}_V\rightarrow\hat{c}_{V\mathrm{P}2}$ with the result
\begin{equation}\label{PI}
	n_{\mathrm{P}2} = \frac{5(1+2\Theta)^2}{3+15\Theta(1+\Theta)}\, .
\end{equation}

Clearly, $n_{\mathrm{P}2} \sim 5/3$ as $\Theta\rightarrow 0$, while $n_{\mathrm{P}2}\sim 4/3$ as $\Theta\rightarrow \infty$; in accordance with non-relativistic and ultra-relativistic limit for degenerate fermionic gas.  

\section{Thermal capacity during virial collapse}

In this section we study a quasi-static sequence of self-gravitating systems in the Newtonian virial regime. The system is assumed to remain close to hydrostatic equilibrium during the evolution, so that each state along the sequence satisfies the virial relation derived in Appendix~3. The gravitational field is not treated as a thermal reservoir; rather, it supplies the binding energy and performs mechanical work during contraction or expansion. The self-gravitating system is described by the CEOS introduced in Sec.~II, while gravity enters through the Newtonian potential energy $\Omega$. This approximation is appropriate as long as the compactness remains small, $GM/(Rc^2)\ll1$, and the local thermodynamic properties of the matter can be treated in a weak-field, locally Minkowskian approximation. In this regime the relativistic effects included in the model refer to the kinetic EOS of the matter, encoded by the parameter $\Theta$, not to a fully relativistic treatment of the gravitational field. The approximation is expected to break down for strongly compact systems, where a Tolman--Oppenheimer--Volkoff treatment and redshifted temperature profiles would be required.

Due to the first law of thermodynamics the change of the system's total energy is
\begin{equation}
	dE = \delta Q + \delta W + d\Omega\, ,
\end{equation}
where $\Omega$ is gravitational potential energy.

The exact hydrostatic virial relation is the pressure-integral relation reviewed in Appendix~3. For the two-component surrogate, Appendix~4 constructs the corresponding one-zone pressure--energy coefficient $\alpha_{\rm sur}(T,V)$ from the total surrogate pressure and internal energy. In the nonrelativistic asymptotic regime $n\rightarrow5/3$, this coefficient approaches $\alpha_{\rm sur}=2$. We use this limiting value in the illustrative nonrelativistic virial path below, so that $\Omega=-2U$, where $U$ is the nongravitational internal energy of the surrogate model. Since the gravitational field interacts with the system through mechanical work, along this path $\delta W=-d\Omega$. Therefore
\begin{equation}\label{work}
	\delta W = -p dV = 2dU\, .
\end{equation}

During the virial collapse energy is lost with the rate $dE=d\Omega/2$.
Ignoring outside heat sources, the heat exchanged along the
nonrelativistic virial path is therefore

\begin{equation}
	\delta Q = - dU \, .
\end{equation}

Hence, the thermal capacity along this virial path is

\begin{equation}
	C_\mathrm{vir}=\frac{\delta Q}{dT} = - \frac{dU}{dT}\, .
\end{equation}
Continuing with the differential of internal energy in the form
\begin{equation}
	dU = C_V dT + \pi_T dV\, , 
\end{equation}
where we express $dV$ from Eq.~\eqref{work} and find 
\begin{equation}
	dU = C_V dT -\frac{2}{p} \pi_T dU\, . 
\end{equation}
Finally we obtain 
\begin{equation}\label{cvir}
	C_\mathrm{vir} = \frac{C_V}{1-\frac{2T}{p}\left(\frac{\partial p}{\partial T}\right)_V}\, .
\end{equation}
Using Eqs.~\eqref{eq:dUdV-surrogate} and \eqref{eq:internal-pressure-model}, Eq.~\eqref{cvir} can be rewritten in the compact form
\begin{equation}\label{cvircompact}
	C_\mathrm{vir} = \frac{p C_V}{2\mathcal{K}\rho^{n(T)}-p}\, .
\end{equation}
The virial heat capacity diverges at, and changes sign across, the pole defined by
\begin{equation}\label{DC}
	p=2\mathcal{K}\rho^{n(T)}\equiv -2\pi_T\, .
\end{equation}
Physical meaning of this condition becomes clear when we consider case with $n$ being constant, then 
\begin{equation}
	p_\mathrm{supp}= \mathcal{K}\rho^{n}\, , 
\end{equation}	
and the total pressure is $p=p_\mathrm{th}+p_\mathrm{supp}$. Hence the divergence condition Eq. (\ref{DC}) becomes
\begin{equation}
	p_\mathrm{th}=p_\mathrm{supp}\, .
\end{equation}
In the support-dominated limit, \(p_{\rm supp}\gg p_{\rm th}\), Eq.~\eqref{cvircompact} gives \(C_{\rm vir}\rightarrow C_V>0\), whereas in the thermal-pressure-dominated limit, \(p_{\rm th}\gg p_{\rm supp}\), one obtains \(C_{\rm vir}\rightarrow -C_V<0\).

\subsection{Relativistic corrections}

The derivation above uses $\alpha_{\rm vir}=2$, corresponding to the nonrelativistic asymptotic limit of the surrogate pressure--energy coefficient derived in Appendix~4. A temperature-only relativistic refinement can be parametrized by
\begin{equation}
	\Omega=-\alpha_{\rm vir}(\Theta)U\, ,
\end{equation}
where $\alpha_{\rm vir}(\Theta)$ interpolates between $2$ in the nonrelativistic limit and $1$ in the ultrarelativistic limit. Equation~\eqref{cvir} is recovered in the nonrelativistic limit.

For a one-component relativistic ideal gas, Appendix~3 gives the explicit temperature-dependent coefficient $\alpha_{\rm J}(\Theta)$. For the two-component surrogate, Appendix~4 constructs the more general one-zone coefficient $\alpha_{\rm sur}(T,V)$ directly from the total surrogate pressure and internal energy; its value also depends on the relative energetic weights of the thermal and support contributions.

 If the temperature dependence of $\alpha_{\rm vir}(\Theta)$ is retained,
 the virial relation gives
 
 \begin{equation}
 	d\Omega
 	=
 	-\alpha_{\rm vir} dU
 	-
 	U d\alpha_{\rm vir}\, ,
 \end{equation}
 
 and hence, along the virial path,
 
 \begin{equation}
 	\delta W
 	=
 	-p dV
 	=
 	-d\Omega
 	=
 	\alpha_{\rm vir}dU
 	+
 	U d\alpha_{\rm vir}\, .
 \end{equation}
 
 Combining this with
 
 \begin{equation}
 	dU=C_VdT+\pi_TdV
 \end{equation}
 
 yields
 
 \begin{equation}
 	\frac{dU}{dT}
 	=
 	\frac{
 		C_V-\dfrac{\pi_T}{p}U\dfrac{d\alpha_{\rm vir}}{dT}
 	}{
 		1+\alpha_{\rm vir}\dfrac{\pi_T}{p}
 	}\, .
 \end{equation}
 
 Furthermore,
 
 \begin{equation}
 	\delta Q
 	=
 	dU-\delta W
 	=
 	(1-\alpha_{\rm vir})dU-U d\alpha_{\rm vir}\, .
 \end{equation}
 
 Using the thermodynamic identity, Eq.~\eqref{eq:internal-pressure-model}, we obtain the
 relativistically corrected virial heat capacity 
 \begin{equation}\label{cvhc}
 	C_{\rm vir}^{(\alpha)}
 	=
 	\frac{
 		[\alpha_{\rm vir}(\Theta)-1]C_V
 		+
 		U\dfrac{d\alpha_{\rm vir}}{dT}
 		\dfrac{T}{p}
 		\left(\dfrac{\partial p}{\partial T}\right)_V
 	}{
 		\alpha_{\rm vir}(\Theta)-1
 		-
 		\alpha_{\rm vir}(\Theta)
 		\dfrac{T}{p}
 		\left(\dfrac{\partial p}{\partial T}\right)_V
 	}\, .
 \end{equation}
 
 For $\alpha_{\rm vir}=2$ and $d\alpha_{\rm vir}/dT=0$, this reduces to
 Eq.~\eqref{cvir}. The divergence condition is therefore shifted to
 
 \begin{equation}
 	p=
 	\frac{\alpha_{\rm vir}(\Theta)}{\alpha_{\rm vir}(\Theta)-1}
 	T\left(\frac{\partial p}{\partial T}\right)_V\, .
 \end{equation}
 
 The additional term proportional to $d\alpha_{\rm vir}/dT$ affects the
 finite value of $C_{\rm vir}^{(\alpha)}$ away from the pole, but it does
 not change the location of the divergence. Thus the relativistic virial
 correction changes the quantitative position of the transition curve,
 while preserving the qualitative interpretation: the sign change of the
 virial heat capacity is still controlled by the competition between
 thermal pressure and the additional stabilizing pressure support.

\subsection{Model}

In the illustrative model below we use the nonrelativistic global virial
coefficient $\alpha_{\rm vir}=2$ and therefore the nonrelativistic
virial-capacity formula, Eq.~\eqref{cvir}. This choice keeps the analytic
surrogate simple and isolates the basic pressure-competition mechanism.
The relativistically corrected expression, Eq.~\eqref{cvhc}, changes the
quantitative position of the transition curve; however, it does not
change the qualitative message of the model: the heat capacity changes
sign when the stabilizing support becomes competitive with the thermal
pressure.

To find the Padé-approximated $C_\mathrm{V}$ entering the
nonrelativistic virial-capacity formula for our two-component
surrogate, we combine caloric EOS postulated in Eq.~\eqref{CEOS} with effective polytropic index given by Eq.~\eqref{PI} and then take derivative with respect to $T$. The resulting relation is
\begin{equation}
	C_{V}\sim \frac{3}{2}Nk_\mathrm{B} + \frac{\mathcal{K}\rho^{n}n^\prime\left[(n-1)\ln\rho-1\right]}{\left(n-1\right)^2}V\, ,
\end{equation} 
where
\begin{equation}
	n^\prime \equiv\frac{\rm{d}n}{\rm{d}T} = -\frac{k_\mathrm{B}}{mc^2}\frac{5(1+2\Theta)}{3\left[1+5\Theta(1+\Theta)\right]^2}\, .
\end{equation}

\begin{figure}[t]
	\begin{tikzpicture}[thick, scale=1]
		
		\node[inner sep=1pt] (img1) at (-0.5,0) {\includegraphics[width=0.45\textwidth]{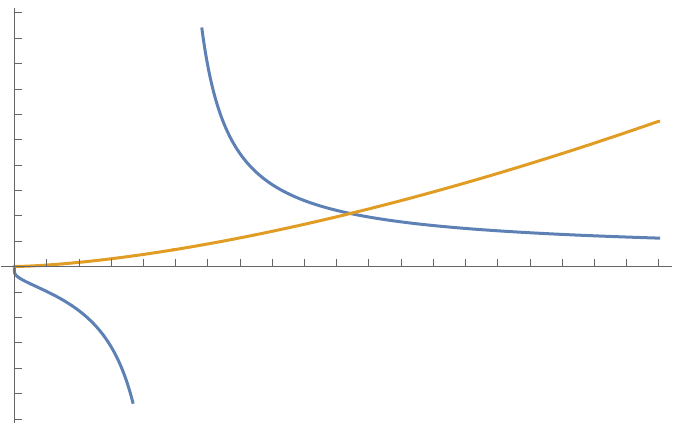}};
		\node at (0,-2.7) {\large $\rho$};
		\node at (1.95,1.1) {\large $p$};
		\node at (-1.2,1.2) {\large $\hat{c}_\mathrm{virP}$};
	\end{tikzpicture}
	\caption{\label{S4} Illustration figure of $\hat{c}_{\mathrm{virP}}$ in dimensionless units as a function of $\rho$ and constant temperature $T$.}	
\end{figure}

To give an order-of-magnitude interpretation of the parameter range, recall that for electrons
\begin{equation}
	mc^2\simeq 511\,{\rm keV}.
\end{equation}
Thus $\Theta=0.1$ corresponds to $k_{\mathrm B}T\simeq 51\,{\rm keV}$, or $T\simeq 6\times10^8\,{\rm K}$. This is already a regime where relativistic corrections to the electron component are non-negligible, while the baryonic component may still be treated as nonrelativistic. The transition region in our surrogate model should therefore be interpreted as an order-of-magnitude representation of the onset of relativistic softening in the stabilizing support sector, rather than as a precise stellar-structure prediction. A quantitatively accurate treatment of a particular compact object would require a microscopic Fermi--Dirac EOS and a full hydrostatic model; here the purpose is to isolate the pressure-competition mechanism controlling the sign of the virial heat capacity.

In Fig.~\ref{S4} we see an illustrative example of the $\hat{c}_\mathrm{virP}(\rho,\Theta)$. Capacity changes sign when the support pressure stabilizes the system.

\section{Schwarzschild Anti-de-Sitter Thermodynamics}

\begin{figure}[t]
	\begin{tikzpicture}[thick, scale=1]
		
		\node[inner sep=1pt] (img1) at (-0.5,0) {\includegraphics[width=0.45\textwidth]{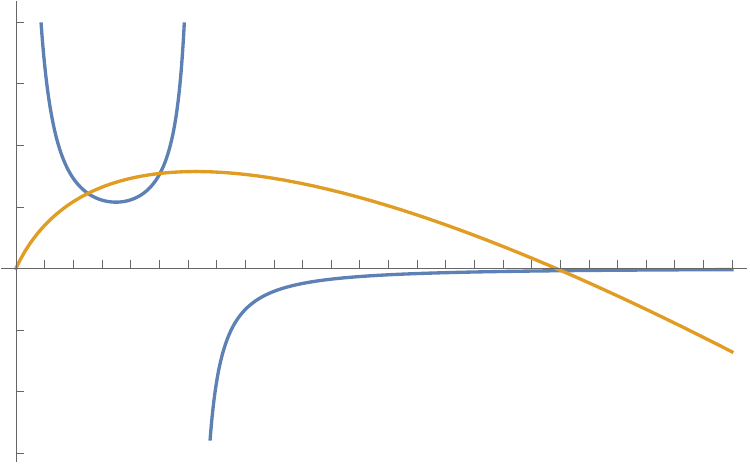}};
		\node at (0,-2.7) {\large $\rho$};
		\node at (-2.3,-1) {\large $\hat{c}_p$};
		\node at (0.2,0.5) {\large $p$};
		\node at (-3.3,2.1) {Big BHs};
		\node at (0.4,-1) {Small BHs};
		\node at (1.1,2.2) {AdS};
		\node at (1.9,2.2) {dS};
		\draw[dash dot,line width=0.2mm] (1.55,-2.4) -- (1.55,2.4);
	\end{tikzpicture}
	\caption{\label{S5} Illustration figure of Schwarzschild AdS BH isobaric heat capacity $\hat{c}_{p}$ and pressure in dimensionless units as a function of $\rho$ with some fixed constant value of temperature. The region of negative pressure loosely corresponds to the Schwarzschild-de Sitter space.}	
\end{figure}

In this section we compare the thermal capacity of the self-gravitating system $C_{\mathrm{virP}}$ with the thermal capacity of the BH.
Particularly, it is convenient to consider thermodynamic properties of the Schwarzschild AdS BH because they allow one to formally define notion of the pressure $p$ and volume $V$ via BH thermodynamics \cite{Bar:Car:Haw:1973:} sub-branch called \textit{black hole chemistry} \cite{Henn:Tei:1984:,Tei:1985:,Kub:Man:2015:}.

The Schwarzschild AdS metric is defined via line element (in Planck units $G=\hbar=c=k_\mathrm{B}=1$)
\begin{equation}
	ds^2 = -fdt^2 + \frac{dr^2}{f} + r^2 d\Omega^2_2\, ,\quad f = 1-\frac{2M}{r} - \frac{\Lambda r^2}{3}\, ,
\end{equation}
with a negative cosmological constant $\left(\Lambda<0\right)$ connected to the pressure $p$ by relation
\begin{equation}
	p = -\frac{\Lambda}{8\pi}\, .
\end{equation} 
The BH horizon lies on $r=r_+$ defined via relation $f(r_+) = 0$. This yields 
\begin{equation}
	M = \frac{r_+}{2}\left(1+\frac{8\pi p r^2_+}{3}\right)\, .
\end{equation}
Here the black-hole mass $M$ is thermodynamically interpreted as enthalpy \cite{Kas:etal:2009:}. With standard Bekenstein--Hawking \cite{Bek:1973:,Haw:1975:} definition of BH entropy, i.e, 
\begin{equation}
	S = \frac{A}{4} = \pi r_+^2\, ,
\end{equation}
where $A$ is the area of the event horizon, we find that thermodynamic volume of the BH is \cite{Kub:Man:Teo:2017:}
\begin{equation}
	V = \left(\frac{\partial M}{\partial p}\right)_S \equiv \left(\frac{\partial M}{\partial p}\right)_{r_+}=\frac{4}{3}\pi r_{+}^3\, .
\end{equation}

The BH temperature is given by the surface gravity $\kappa = f^\prime(r_+)/2$ as
\begin{equation}\label{Tp}
	T = \frac{\kappa}{2\pi} = \frac{1}{4\pi}\left.\frac{df(r)}{dr}\right|_{r=r_+} = \frac{1+8\pi pr^2_+}{4\pi r_+}\, .
\end{equation}
It is useful to express these relations in natural variables for enthalpy, i.e., as functions of $S$ and $p$
\begin{eqnarray}
	M &=& \frac{2+\zeta}{6}\sqrt{\frac{S}{\pi}}\, ,\quad T = \frac{\zeta}{4\sqrt{\pi S}}\, ,\quad V = \frac{4}{3}\sqrt{\frac{S^3}{\pi}}\, ,\nonumber
\end{eqnarray}
where we have provided useful substitution $\zeta = 1 + 8pS$. 

Thermal isobaric ($\Lambda$ = const.) capacity is 
\begin{equation}\label{capp}
	C_p = \left(\frac{\partial M}{\partial T}\right)_p = -\frac{2S\zeta}{1-8pS}\, .
\end{equation}
and usually diverges at so-called Davies critical points \cite{Dav:1977:}. In our case of Schwarzschild AdS BH we have one critical point at
\begin{equation}
p = \frac{1}{8S}\, .
\end{equation} 
Since the pressure can be expressed from Eq.~\eqref{Tp} in the form
\begin{equation}
	p = \frac{T}{2r_+}-\frac{1}{8\pi r_+^2}
\end{equation} 
we can easily check, via Eq.~\eqref{eq:internal-pressure-model}, that at critical point the pressure is equal the internal pressure, i.e., 
\begin{equation}
	p = \frac{1}{8S} = \pi_T\, .
\end{equation} 
Davies critical points are often interpreted as indicators of second-order phase transitions (in the appropriate thermodynamic ensemble \cite{Cur:1981:}).

Putting Eq.~\eqref{capp} into general Mayer's relation, i.e,
\begin{equation}
	C_p = C_V + \left[\vphantom{\int}p+\pi_T\right]\left(\frac{\partial V}{\partial T}\right)_p 
\end{equation}
we find that isochoric heat capacity of Schwarzschild AdS BH, $C_V = 0$. This is no surprise. It is caused by the degeneracy of the Legendre transformation between BH enthalpy $H\equiv M$ and its internal energy $U$. We find that $U = M-pV\equiv U(V)$ is only a function of $V$ and does not depend on $T$. This is rather opposite behavior than of the ideal gas. So to compare $C_\mathrm{vir}$ to BH $C_V$ is pointless, but we can compare  $C_\mathrm{vir}$ to $C_p$.  

To that goal, let us formally define average density of the BH as 
\begin{equation}
	\rho \equiv \frac{M}{V} = p+\frac{3}{8S}\, .
\end{equation}
Then we can express isobaric heat capacity (\ref{capp}) and vacuum pressure as a function of variables $\rho$ and $T$.  

In Fig.~\ref{S5} we can see pressure's typical dependence on $\rho$ and also behavior dimensionless isobaric heat capacity $\hat{c}_\mathrm{p}$ for constant temperature $T$. We see the presence of second-order phase transition between small and large BHs \cite{Alt:Kub:Mann:2013:}.  

Interestingly, according to Eq.~\eqref{capp}, the phase transition occurs at point $p=\pi_T$, where BH pressure equals internal pressure. Also, in Fig.~\ref{S5} it is suggested that the point $p=\pi_T$ is local maximum of the function  $p(\rho,T)$. It can be easily checked that it is indeed so. 

We can also see that our analysis formally goes beyond AdS spaces into the dS spaces with negative pressure. But in such spaces there is also cosmological horizon which complicates precise analysis. For small negative pressures our analysis should be quite precise, but as the pressure goes further down, it no longer can be applied. For de Sitter space the similar analysis was done in \cite{Bar:2024:} with the aid of Newman--Penrose scalars \cite{New:Pen:1962:}.   

This comparison shows similar heat-capacity behavior in the self-gravitating system and the Schwarzschild AdS BH.

\section{Conclusions}

In this work, we have constructed a relativistic two-component model of a self-gravitating system that mimics the thermodynamic behavior of partially degenerate matter via a temperature-dependent heat capacity. By combining  an ideal gas component with effective barotropic support we achieved a phase transition between regimes of negative and positive heat capacity, capturing key features of gravitational collapse and thermal stabilization.

We employed matched asymptotic expansions and Padé approximants to express the specific heat capacity $c_V$ in a globally accurate analytic form, allowing for tractable analysis without sacrificing physical fidelity. The  model was then compared to the thermodynamics of Schwarzschild--Anti-de Sitter black holes using the extended phase space framework of black hole chemistry. Remarkably, both systems exhibit transitions in heat capacity behavior tied to internal pressure balances: degeneracy pressure in the case of the gas, and AdS vacuum pressure in the black hole.

Our findings reinforce the conceptual analogy between black holes and relativistic self-gravitating systems, highlighting the shared thermodynamic structures that emerge from gravitational binding. This may serve as a useful heuristic framework for interpreting black hole heat capacities and their relation to classical astrophysical systems.

Future work could extend this approach to rotating black holes, charged configurations, or systems with true quantum degeneracy to explore how far this thermodynamic correspondence can be generalized.

\paragraph*{Outlook.}
Further extensions of the present framework include (i) rotation and/or electric charge, where accretion-driven evolution can lead to qualitatively different final states (including exotic compact-object scenarios beyond the standard Kerr--Newman family), and (ii) additional sources of support such as external magnetic fields and charged fluids, which can drastically modify equilibrium configurations (e.g. non-conducting tori around magnetized black holes).
For representative studies in these directions see, e.g., \cite{Blaschke2022,Stuchlk2022}.

\section*{Acknowledgements}

The work was supported by Research Centre for Relativistic Physics and Astrophysics, Institute of Physics, Silesian University in Opava.

\section*{References}
\bibliography{ref}
\newpage
\section*{Appendix 1 - Solution to integral}

We would like to find solution to one-particle partition function given by Eq.~\eqref{OPF}. Thus, we have to solve the integral
\begin{equation}
	I = \mathrm{e}^{\beta mc^2}\int\limits_{0}^{\infty} p^2 \mathrm{e}^{-\beta\sqrt{p^2 c^2 + m^2 c^4}}dp\, . 
\end{equation}
After substituting $p=mc \sinh(t)$ we get 
\begin{eqnarray}
	I &=& m^3 c^3 \mathrm{e}^{\beta mc^2}\int\limits_{0}^{\infty} \mathrm{e}^{-\beta mc^2 \cosh(t)}\sinh^2(t)\cosh(t)dt\nonumber \\
	&=& m^3 c^3 \mathrm{e}^{\beta mc^2}\int\limits_{0}^{\infty} \mathrm{e}^{-\beta mc^2 \cosh(t)}\cosh^3(t)dt\nonumber \\
	&-& m^3 c^3 \mathrm{e}^{\beta mc^2}\int\limits_{0}^{\infty} \mathrm{e}^{-\beta mc^2 \cosh(t)}\cosh(t)dt\, , 
\end{eqnarray}	
where we have used $\sinh^2(t)=\cosh^2(t)-1$. 

Next we introduce the modified Bessel functions of the second kind which are defined via  \cite{NIST}
\begin{equation}
	K_\nu(x) = \int\limits_{0}^{\infty} \mathrm{e}^{-x \cosh(t)}\cosh(\nu t)dt
\end{equation}
Therefore we have
\begin{equation}
	I = m^3 c^3 \mathrm{e}^{1/\Theta}\left[\frac{d^2}{d (1/\Theta)^2}K_1(1/\Theta)-K_1(1/\Theta)\right]\, ,
\end{equation}
where the relativistic parameter $\Theta$ is given by Eq. (\ref{relpar}).
 
With the aid of the identity
\begin{equation}
	\frac{d^2}{dx^2}K_1(x)-K_1(x)=\frac{K_2(x)}{x}\, ,
\end{equation}
we find one-particle partition function in the form
\begin{equation}
	z_1 = \frac{Vm^3c^3}{2\pi^2\hbar^3}\mathrm{e}^{1/\Theta}\Theta K_2\, ,
\end{equation}
where $K_2$ is a function of $1/\Theta$.

\section*{Appendix 2 - Properties of relativistic ideal  gas}

Relativistic ideal gas is sometimes called J{\" u}ttner gas \cite{Juttner1911}.

Partition function for the whole system is
\begin{equation}
	z = \frac{z_1^N}{N!}\, .
\end{equation}

Using the canonical identity $F=-k_{\mathrm B}T\ln z$ (and Stirling’s approximation for $N!$), we obtain Helmholtz free energy of J{\" u}ttner gas in the form
\begin{equation}
	F = -N mc^2 \left[\Theta+1+\Theta\ln\left(\frac{\alpha V\Theta K_2}{N}\right)\right]\, ,
\end{equation} 
with 
\begin{equation}
	\alpha \equiv \frac{m^3c^3}{2\pi^2\hbar^3}\, .
\end{equation}

Note that using the reduced Compton wavelength $\bar\lambda_C=\hbar/(mc)$, the prefactor
\[
\alpha=\frac{1}{2\pi^{2}\,\bar\lambda_C^{\,3}}
\]
sets a Compton phase–space density scale. In particular, the
combination $\alpha\,e^{1/\Theta}\,\Theta\,K_{2}(1/\Theta)$ that appears in $z_1$
reproduces the nonrelativistic limit via the thermal de Broglie wavelength
\[
\lambda_T=\frac{h}{\sqrt{2\pi m k_{\mathrm B} T}}
\qquad\Longrightarrow\qquad
\frac{1}{\lambda_T^{3}}
=\Big(\frac{m k_{\mathrm B} T}{2\pi \hbar^{2}}\Big)^{3/2}.
\]
If the spin degeneracy $g$ is included, $\alpha$ should be multiplied by $g$.

Gibbs--Helmholtz equation then gives
\begin{equation}\label{UJ}
	U =-T^2\left(\frac{\partial}{\partial T}\right)_V\frac{F}{T} = N mc^2 \left(\frac{K_3}{K_2}-\Theta-1\right)\, .
\end{equation}
Equation of state is given by
\begin{equation}\label{EOS1}
	p = -\left(\frac{\partial F}{\partial V}\right)_T = \frac{Nk_{\mathrm B}T}{V}\, .
\end{equation}
And finally, specific and dimensionless isochoric thermal capacity, i.e,  
\begin{equation}
	\hat{c}_V \equiv \frac{1}{Nk_{\mathrm B}}\left(\frac{\partial U}{\partial T}\right)_V
\end{equation}
is 
\begin{equation}
	\hat{c}_V = \left(1+\frac{1}{2\Theta^2}-\frac{\left(K_1+K_3\right)^2-K_2\left(K_0+K_4\right)}{4K_2^2}\right)\, .\nonumber
\end{equation}
With well known identities between Bessel functions this simplifies into  
\begin{equation}
\hat c_V
= 3 + \frac{1}{\Theta^{2}}
- \frac{3}{\Theta}\,\frac{K_{1}}{K_{2}}
- \frac{1}{\Theta^{2}}\!\left(\frac{K_{1}}{K_{2}}\right)^{\!2}.
\end{equation}

\section*{APPENDIX 3 - VIRIAL THEOREM AND RELATIVISTIC INTERPOLATION}

The pressure--energy coefficient entering the virial relation is fixed by
the thermodynamic properties of the system. We first derive the
Newtonian pressure virial theorem and then evaluate the pressure--energy
coefficient for a one-component relativistic ideal gas, whose J\"uttner
EOS gives an explicit interpolation between the familiar
nonrelativistic and ultrarelativistic limits. The corresponding
coefficient of the two-component surrogate is constructed from its total
pressure and internal energy in Appendix~4 (for a more in-depth analysis
of the virial theorem in weak gravity see, e.g. \cite{Pad:2002:}).

The derivation starts from the Newtonian hydrostatic equilibrium equation
\begin{equation}
	\nabla p=-\varrho \nabla \phi ,
	\label{eq:app-hydrostatic-equilibrium}
\end{equation}
where $p$ is the local pressure, $\varrho$ is the mass density, and
$\phi$ is the Newtonian gravitational potential. Multiplying
Eq.~\eqref{eq:app-hydrostatic-equilibrium} by the position vector
$\mathbf r$ and integrating over the volume gives
\begin{equation}
	\int_V \mathbf r\cdot \nabla p\,d^3x
	=
	-
	\int_V \varrho\, \mathbf r\cdot \nabla \phi\,d^3x .
	\label{eq:app-virial-start}
\end{equation}
Using Gauss' theorem, the left-hand side becomes
\begin{equation}
	3p_sV-3\int_V p\,d^3x ,
	\label{eq:app-gauss-pressure}
\end{equation}
where $p_s$ is the surface pressure. For Newtonian gravity the
right-hand side of Eq.~\eqref{eq:app-virial-start} is the gravitational
potential energy $\Omega$. Hence one obtains the scalar virial theorem
\begin{equation}
	\Omega+3\int_V p\,d^3x=3p_sV .
	\label{eq:app-hydrostatic-virial}
\end{equation}
For negligible surface pressure this reduces to
\begin{equation}
	\Omega+3\int_V p\,d^3x=0 .
	\label{eq:app-zero-surface-virial}
\end{equation}
This is the form of the virial theorem appropriate for a
self-gravitating system in hydrostatic equilibrium; see, e.g.,
Refs.~\cite{Chandrasekhar1939,BinneyTremaine2008}.

	To rewrite Eq.~\eqref{eq:app-zero-surface-virial} as a relation between
	$\Omega$ and $U$, we must specify how the pressure is
	related to the local energy density $u$. For a nonrelativistic monatomic
	ideal gas we have

\begin{equation}
		u=\frac{3}{2}p ,
	\label{eq:app-nr-caloric-density}
\end{equation}

	For

\begin{equation}
		U=\int_V u\,d^3x .
	\label{eq:app-kinetic-energy}
\end{equation}

	the Eq.~\eqref{eq:app-zero-surface-virial} gives

\begin{equation}
		2U+\Omega=0 .
	\label{eq:app-nr-virial}
\end{equation}

	This is the standard nonrelativistic virial theorem.
	
	For an ultrarelativistic ideal gas one has instead

\begin{equation}
		u=3p ,
	\label{eq:app-ur-caloric-density}
\end{equation}

	and therefore

\begin{equation}
		U+\Omega=0 .
	\label{eq:app-ur-virial}
\end{equation}

	Thus the coefficient multiplying the kinetic internal energy changes
	from $2$ in the nonrelativistic limit to $1$ in the ultrarelativistic
	limit.

	The relativistic interpolation between these two limits is obtained from
	the J\"uttner gas with the internal energy density above rest mass (see Eq. (\ref{UJ}) in appendix 2) in the form

\begin{equation}
		u=n_{\rm p}mc^2
		\left[
		\frac{K_3}{K_2}
		-\Theta-1
		\right],
	\label{eq:app-juttner-energy-density}
\end{equation}

	where $n_{\rm p}$ is the particle-number density, $\Theta$ is the relativistic parameter given by Eq. (\ref{relpar}) and
	$K_\nu\equiv K_\nu(1/\Theta)$ denotes the modified Bessel function of
	the second kind. The ideal-gas pressure is
	$p=n_{\rm p}k_{\rm B}T=n_{\rm p}mc^2\Theta$. Hence

\begin{equation}
		p
		=
		\frac{\alpha_{\rm J}(\Theta)}{3}u,
	\label{eq:app-juttner-pressure-energy}
\end{equation}

	with

\begin{equation}
		\alpha_{\rm J}(\Theta)
		=
		\frac{3\Theta}
		{
			K_3/K_2-\Theta-1
		} .
	\label{eq:app-alpha-juttner}
\end{equation}

	This expression has the correct limiting behavior,

\[
		\alpha_{\rm J}(\Theta)\rightarrow 2
		\quad
		(\Theta\rightarrow0),
		\qquad
		\alpha_{\rm J}(\Theta)\rightarrow 1
		\quad
		(\Theta\rightarrow\infty).
	\label{eq:app-alpha-limits}
\]

	For a quasi-static global configuration characterized by one effective
	temperature parameter $\Theta$, the pressure integral is

\begin{equation}
		3\int_V p\,d^3x
		=
		\alpha_{\rm J}(\Theta)U.
	\label{eq:app-juttner-global-pressure}
\end{equation}

	Substitution into Eq.~\eqref{eq:app-zero-surface-virial} gives

\begin{equation}
		\alpha_{\rm J}(\Theta)U+\Omega=0 .
	\label{eq:app-juttner-global-virial}
\end{equation}

	For practical analytic work we use a compact Pad\'e approximation to the
	J\"uttner coefficient. Expanding Eq.~\eqref{eq:app-alpha-juttner} at
	small $\Theta$ gives

\begin{equation}
		\alpha_{\rm J}(\Theta)
		=
		2-\frac{5}{2}\Theta+\frac{45}{8}\Theta^2
		+O(\Theta^3).
	\label{eq:app-alpha-expansion}
\end{equation}

	A simple two-point Pad\'e expression that reproduces this expansion up to
	order $\Theta^2$ and has the correct ultrarelativistic limit is

\begin{equation}
		\alpha_{P2}(\Theta)
		=
		\frac{
			2+5\Theta+\frac{15}{4}\Theta^2
		}{
			1+\frac{15}{4}\Theta+\frac{15}{4}\Theta^2
		} .
	\label{eq:app-alpha-pade}
\end{equation}

The Padé-improved Jüttner relation can therefore be written as
\begin{equation}
		\alpha_{P2}(\Theta)U+\Omega=0\, .
	\label{eq:app-pade-virial}
\end{equation}
Here $\alpha_{P2}(\Theta)$ parametrizes the pressure--energy relation of
the one-component J\"uttner gas. The effective coefficient
$\alpha_{\rm sur}(T,V)$ of the two-component surrogate is obtained from
its total $p(T,V)$ and $U(T,V)$ in Appendix~4.

For the detailed behavior of $\alpha_{\rm J}(\Theta)$ and $\alpha_{P2}(\Theta)$, see Fig.~\ref{AS1}; Fig.~\ref{AS11} shows a magnified view near the transition region.

\begin{widetext}
	\begin{minipage}{0.48\linewidth}
		\begin{tikzpicture}[thick, scale=1]
			\node[rotate=-6] at (2.3,-1.05) {$\alpha_{\mathrm{J}}$};	
			\node[inner sep=1pt] (img1) at (-0.5,0) {\includegraphics[width=0.9\textwidth]{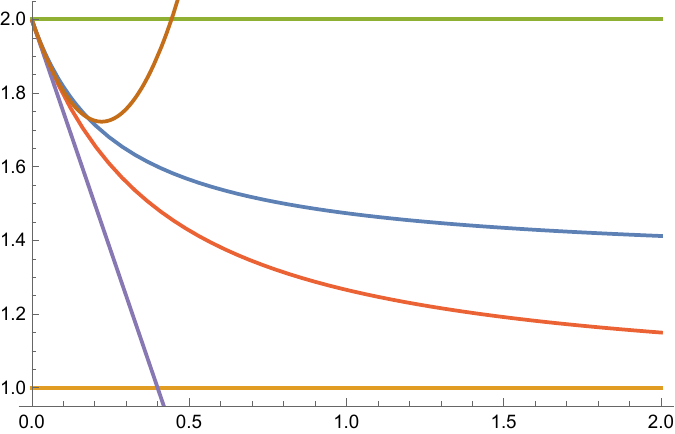}};
			\node at (0,-2.7) {\large $\Theta$};
			\node at (-4.7,0) {\large $\alpha_\mathrm{J}$};
			\node at (2.1,-1.75) {nonrelativistic};
			\node at (2.0,2.45) {ultra relativistic};
			\node[rotate=-2] at (2.1,0.05) {$\alpha_{P1}$};
			\node[rotate=-72] at (-3.3,-0.65) {Taylor, 1. order};
			\node[rotate=0] at (-1.85,1.75) {2. order};	
			\draw[dashed,gray] (-0.6,-0.95) rectangle (-0.2,-0.75);
			
		\end{tikzpicture}
		\captionof{figure}{\label{AS1} Comparison of the function $\alpha_\mathrm{J}(\Theta)$  at five different levels of accuracy. Matched
			asymptotic expansions and exact Bessel form  merge with ultra relativistic case at $\Theta\rightarrow \infty$.}
	\end{minipage}
	\hfill
	\begin{minipage}{0.48\linewidth}
		\begin{tikzpicture}[thick, scale=1]
			
			\node[inner sep=1pt] (img2) at (-0.5,0) {\includegraphics[width=0.9\textwidth]{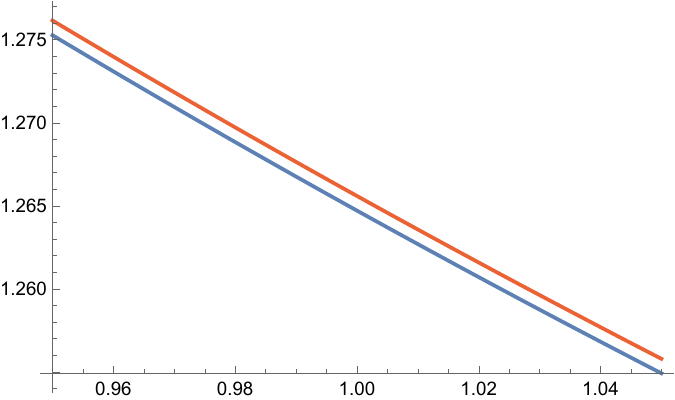}};
			\node at (0,-2.7) {\large $\Theta$};
			\node at (-4.7,0) {\large $\alpha_\mathrm{J}$};
			\node[rotate=-28] at (0.4,-0.1) {$\alpha_\mathrm{J}$};	
			\node[rotate=-28] at (0.4,-0.8) {$\alpha_{P2}$};	
		\end{tikzpicture}
		\captionof{figure}{\label{AS11} Zoom of Fig.~\ref{AS1}, showing the accuracy of the second-order Padé approximant near the transition region.}
	\end{minipage}
\end{widetext}

\section*{APPENDIX 4 - EFFECTIVE PRESSURE--ENERGY COEFFICIENT OF THE SURROGATE MODEL}

For the two-component surrogate, the effective pressure--energy
coefficient follows from the total surrogate pressure $p(T,V)$ and total
nongravitational internal energy $U(T,V)$. The hydrostatic starting point
is the pressure virial theorem, Eq.~\eqref{eq:app-zero-surface-virial},
with $p$ denoting the total pressure. In the one-zone description used in
the main text, $T$, $V$, $\varrho$, and $p$ are effective global
quantities, and $p(T,V)$ represents the volume-averaged pressure entering
the virial integral,
\begin{equation}
	\int_{V} p(\mathbf r)\,d^3x
	\simeq p(T,V)V .
	\label{one_zone_pressure_app4}
\end{equation}
The one-zone virial relation can then be written as
\begin{equation}
	\Omega + 3p(T,V)V = 0 .
	\label{one_zone_virial_app4}
\end{equation}

It is useful to express the factor \(3pV\) in terms of the
nongravitational internal energy \(U\) of the surrogate model. We define
the effective pressure--energy coefficient
\begin{equation}
	3p(T,V)V
	\equiv
	\alpha_{\rm sur}(T,V)U(T,V)\, .
	\label{alpha_sur_def_app4}
\end{equation}

This definition parametrizes the one-zone pressure--energy relation
implied by the caloric equation of state and the pressure relation of the
surrogate model.

For the full interpolation region, where $n(T)$ varies with temperature,
the support contribution entering the total pressure is given by the
integral relation
\begin{equation}
	p_{\rm supp}(T,V)
	=
	\mathcal{K}T\int_T^\infty
	\frac{\varrho^{\,n(\tau)}}{\tau^2}\,d\tau .
	\label{psupp_full_app4}
\end{equation}
Thus the exact expression for $\alpha_{\rm sur}$ is generally a
non-elementary function of the effective temperature and density. In the
asymptotic regimes, where $n(T)$ approaches a constant, the support
pressure reduces to the standard polytropic form and the structure of
$\alpha_{\rm sur}$ becomes explicit.

Let \(n\) be constant. Then
\begin{equation}
	p(T,V)
	=
	p_{\rm th}+p_{\rm supp}
	=
	\frac{Nk_{\rm B}T}{V}
	+
	\mathcal{K}\varrho^n ,
	\label{dalton_app4}
\end{equation}
and the caloric equation of state becomes
\begin{equation}
	U
	\equiv
	U_{\rm th}+U_{\rm supp}
	=
	\frac32 Nk_{\rm B}T
	+
	\frac{\mathcal{K}}{n-1}\varrho^n V .
	\label{U_split_app4}
\end{equation}
Therefore
\begin{equation}
	3pV
	=
	3p_{\rm th}V+3p_{\rm supp}V
	=
	2U_{\rm th}+3(n-1)U_{\rm supp}.
	\label{pressure_energy_split_app4}
\end{equation}
Comparing with Eq.~\eqref{alpha_sur_def_app4}, we obtain
\begin{equation}
	\alpha_{\rm sur}
	=
	\frac{
		2U_{\rm th}+3(n-1)U_{\rm supp}
	}{
		U_{\rm th}+U_{\rm supp}
	} .
	\label{alpha_sur_constant_n_app4}
\end{equation}
Thus the effective coefficient is a weighted pressure--energy coefficient
of the thermal and support sectors.

In the low-temperature asymptotic limit of the surrogate model,
\begin{equation}
	n\to \frac53 ,
\end{equation}
and therefore \(3(n-1)=2\). Equation
\eqref{pressure_energy_split_app4} then gives
\begin{equation}
	3pV = 2U_{\rm th}+2U_{\rm supp}=2U .
	\label{low_T_alpha_app4}
\end{equation}
Hence
\begin{equation}
	\alpha_{\rm sur}\to 2 ,
\end{equation}
independently of which component dominates the internal energy. In this
limit the one-zone virial relation becomes the familiar nonrelativistic
form
\begin{equation}
	\Omega+2U=0 .
\end{equation}

In the high-temperature asymptotic limit,
\begin{equation}
	n\to \frac43 ,
\end{equation}
and therefore \(3(n-1)=1\). Equation
\eqref{pressure_energy_split_app4} gives
\begin{equation}
	3pV = 2U_{\rm th}+U_{\rm supp}.
	\label{high_T_split_app4}
\end{equation}
The corresponding effective coefficient is
\begin{equation}
	\alpha_{\rm sur}
	=
	\frac{2U_{\rm th}+U_{\rm supp}}
	{U_{\rm th}+U_{\rm supp}} .
	\label{alpha_high_T_app4}
\end{equation}
If the support contribution dominates the internal energy,
\(U_{\rm supp}\gg U_{\rm th}\), then
\begin{equation}
	\alpha_{\rm sur}\to 1,
\end{equation}
and the one-zone virial relation approaches the ultrarelativistic form
\begin{equation}
	\Omega+U=0 .
\end{equation}
On the other hand, if \(U_{\rm th}\gg U_{\rm supp}\), even in the
asymptotic regime \(n\to4/3\), one still obtains
\(\alpha_{\rm sur}\to2\). This illustrates that the temperature controls
the effective stiffness of the support sector, while the global virial
coefficient is also determined by the relative energetic weight of the
thermal and support contributions.

The coefficient $\alpha_{\rm sur}$ therefore provides the effective
one-zone pressure--energy coefficient of the two-component surrogate. In
the constant-index asymptotic regimes it follows directly from the
pressure and caloric equations of state. In the full interpolation
region it is a smooth model-dependent function determined by the
integral support-pressure relation, connecting the nonrelativistic and
ultrarelativistic pressure--energy regimes.

\end{document}